\documentclass[prl,twocolumn,showpacs,superscriptaddress,nofootinbib]{revtex4-1}
\usepackage{dcolumn}
\usepackage{bm}
\usepackage{graphicx}
\usepackage{amsfonts}
\usepackage{bbm}
\usepackage{dsfont}
\usepackage{float}
\usepackage{graphicx}
\usepackage{bm}
\usepackage{amssymb}
\usepackage{multibib}
\usepackage{color}
\usepackage{enumerate}
\usepackage{amsmath}
\usepackage{amstext}
\usepackage{latexsym}
\usepackage{braket}
\usepackage{appendix}
\usepackage{textcomp}
\usepackage[usenames,dvipsnames]{xcolor}
\usepackage[colorlinks=true,citecolor=Blue,linkcolor=RubineRed,urlcolor=Blue]{hyperref}

\newcommand{\beq}{\begin{equation}}
\newcommand{\eeq}{\end{equation}}
\newcommand{\beqa}{\begin{eqnarray}}
\newcommand{\eeqa}{\end{eqnarray}}

\begin{document}

\title{Thermofield dynamics: Quantum Chaos versus Decoherence}
\date{\today}
\author{Zhenyu Xu}
\affiliation{School of Physical Science and Technology, Soochow University,
Suzhou 215006, China}
\author{Aurelia Chenu}
\affiliation{Donostia International Physics Center,  E-20018 San
Sebasti\'an, Spain} \affiliation{IKERBASQUE, Basque Foundation for Science,
E-48013 Bilbao, Spain} \affiliation{Massachusetts Institute of Technology, Cambridge, MA 02139, USA}

\author{Toma\v{z} Prosen}
\affiliation{Faculty of Mathematics and Physics, University of Ljubljana, Jadranska ulica 19, 1000 Ljubljana, Slovenia}
\author{Adolfo del Campo}
\affiliation{Donostia International Physics Center,  E-20018 San
Sebasti\'an, Spain} \affiliation{IKERBASQUE, Basque Foundation for Science,
E-48013 Bilbao, Spain} \affiliation{Department of Physics, University of
Massachusetts, Boston, MA 02125, USA}

\begin{abstract}
Quantum chaos imposes universal spectral signatures that govern  the thermofield dynamics of a many-body system in isolation.
The fidelity between the
initial and time-evolving thermofield double states exhibits  as a function of time a decay, dip, ramp and
plateau. Sources of decoherence give rise to a nonunitary evolution
and result in information loss.  Energy dephasing gradually suppresses  quantum noise fluctuations and the dip associated with spectral correlations. Decoherence further delays the appearance of the dip and shortens the span of the linear ramp associated with chaotic behavior.  The interplay between signatures of quantum chaos and information loss is determined by the competition among  the decoherence, dip and plateau characteristic times, as demoonstrated in the stochastic Sachdev-Ye-Kitaev model. \end{abstract}

\maketitle

In an isolated many-body quantum system, quantum chaos imposes universal
spectral signatures such as the form of the eigenvalue spacing distribution. The latter changes from a Poissonian to a Wigner-Dyson distribution as the integrability of the system is broken to make it increasingly chaotic. Such a change in the properties of the system can often be induced, e.g. in many-body spin systems,  by changing a control parameter  \cite{poilblanc1993,Haake,Guhr98,Borgonovi16}.

The Fourier transform of the eigenvalue distribution was soon recognized as a convenient tool to diagnose quantum chaos \cite{Leviandier86,WilkieBrumer91,Alhassid93,Ma95}.
The partition function of the system  analytically continued in the complex-temperature plane has more recently been considered \cite{Cotler2017,Dyer2017,delcampo17}, and it reduces to the former for a purely imaginary inverse temperature $\beta=it$. The quantity $Z(\beta+it)$ is indeed the  complex Fourier transform of the density of states and its absolute square value is related to the spectral form factor. It is also related to the Loschmidt echo \cite{Gorin06,Jacquod09,Bin20} and  quantum work statistics \cite{Chenu2017e,GarciaMata17,Arrais18,chenu2019}.

Complex partition functions take a new meaning in the context of thermofield dynamics \cite{Umezawa82}. Given an equilibrium thermal state of single copy of a quantum system, it is often convenient to consider its purification in an enlarged Hilbert space, which is given by a specific entangled state   between the original and a second copy of the system.  The resulting thermofield double (TFD) state  was recognized early on to be useful in estimating thermal averages of observables \cite{Umezawa82}. The TFD plays also a prominent role in   the description of eternal blackholes and wormholes in AdS/CFT. The fidelity between a given TFD and its time evolution under unitary dynamics is precisely described by the complex Fourier transform of the eigenvalue distribution, specifically, by the absolute square value of the partition function with complex-valued temperature \cite{Papadodimas15,delcampo17}.

Unitarity imposes important constraints on the thermofield dynamics.
The time-evolved state may exhibit highly non-trivial quantum correlations, but the information encoded in the initial state, once scrambled, can in
principle be recovered by reversing the dynamics in an idealized setting. As
a result, the von Neumann entropy of the system remains constant during the
evolution. This feature remains true for the mixed state resulting from averaging over a Hamiltonian ensemble.
The spectral form
factor in an isolated chaotic system displays a decay from unit value leading to a dip, also known as correlation hole, a
subsequent ramp, and a saturation at an asymptotic plateau, in systems
characterized by a finite Hilbert space dimension \cite{Cotler2017,Dyer2017,delcampo17,Schiulaz19}, while its somewhat simpler structure in Floquet many-body systems is
only recently becoming analytically explained \cite{KLP18,BKP18,Chalker18}. Yet, isolated quantum systems
are an idealization. Any realistic quantum system is embedded in a
surrounding environment, the rest of the universe.
Decoherence stems from the interaction between the system
and the surrounding environment, which leads to the build-up of quantum correlations between the
two,  and their entanglement. The environment is generally expected
to be complex and its degrees of freedom unavailable. Information loss in
the system can be traced back to the leakage of information into the
inaccessible environment. The dynamics of the system is non-unitary and its
von Neumann entropy is no longer constant \cite{Zurek03,BreuerBook}.

The interplay between spectral signatures of quantum chaos, decoherence  and information loss is thus a long-standing open problem \cite{Haake,Jalabert01,ZurekPaz94,Karkuszewski02,Can19}. We focus on its role in thermofield dynamics,
with applications ranging from non-equilibrium many-body physics to machine learning
with quantum neural networks in noisy intermediate-scale quantum computers and simulators \cite{ShenZhai20}.
As we shall see, (energy) decoherence washes out short time signatures of quantum chaos, such as the dip in the spectral form factor (correlation hole), while it preserves its long time ramp, conditioned by a competition of characteristic time scales that we elucidate.
As a test-bed, we consider Sachdev-Ye-Kitaev (SYK) model of Majorana fermions involving all-to-all four-body
interactions with quenched disorder  \cite{SachdevYe93,AK15}
which saturates the bound on chaos and admits a gravitational dual,
making it a prominent example \cite{Polchinski16,Maldacena2016}
of holography \cite{Maldacena99}.

\textit{Setting.---}
Consider a system described by a Hamiltonian $H$, with spectral
decomposition in the system Hilbert space $\mathcal{H}$ given by $%
H=\sum_{n=1}^dE_n|n\rangle\langle n|$, $E_n$ being the energy eigenvalues.
A canonical thermal state
of the system at inverse temperature $\beta$ is described by the operator $%
\rho_{\mathrm{th}} = e^{-\beta H} / Z(\beta)$, the partition function of the
system being $Z(\beta)=\mathrm{Tr} (e^{-\beta H})$.
The thermal density matrix can be obtained from a pure, entangled state in
an enlarged Hilbert space $\tilde{\mathcal{H}}=\mathcal{H}\otimes \mathcal{H}
$. Namely, a second copy of the system is used to create the state known as
the thermofield double (TFD) state \cite{Umezawa82} and defined as $|\mathrm{%
TFD}\rangle =\sum_{n}\sqrt{p_n}|nn\rangle$ where $p_{n}=e^{-\beta E_{n}}/Z(\beta )$ and
$\ket{nn}=\ket{n}\otimes \ket{n}$ in $\tilde{\mathcal{H}}$. The
reduced density matrix obtained by tracing over any one of the two copies, $%
\sum_{n}\langle n|\mathrm{TFD\rangle \langle TFD}|n\rangle$, corresponds to
the single-copy canonical thermal state $\rho_{\mathrm{th}}$. Note that the TFD is not invariant under the
unitary $U_{t}=\exp \big[-it(H\otimes \mathds{1}+\mathds{1}\otimes H)\big]$,
taking $\hbar =1$.

The fidelity between the initial TFD state and its evolution provides a measure of quantum chaos \cite%
{Papadodimas15,delcampo17}
\begin{equation}
F_{t}=|\langle \mathrm{TFD}|U_{t}|\mathrm{TFD}\rangle |^{2}=\left\vert \frac{%
Z(\beta +i2t)}{Z(\beta )}\right\vert ^{2}.  \label{FidelityU}
\end{equation}%
In the presence of decoherence, the evolution is not unitary and can
generally be associated with a quantum channel $\Lambda _{t}$ that maps the
initial density matrix to the time-evolved state, i.e., $\rho _{t}=\Lambda
_{t}[\rho _{0}]$. The fidelity between two mixed states $\rho _{0}$ and $%
\rho _{t}$ generalizes the notion of overlap between pure states. It is
defined as $\big(\mathrm{Tr}\sqrt{\sqrt{\rho _{0}}\rho _{t}\sqrt{\rho _{0}}}%
\big)^{2}$ and takes a particularly simple form when the initial state is
pure. We shall thus be interested in the fidelity between the initial (pure)
TFD state and its evolution, i.e.,
\begin{equation}
F_{t}=\langle \mathrm{TFD}|\Lambda _{t}[\tilde{\rho}_{0}]|\mathrm{TFD}%
\rangle ,  \label{Fidelity}
\end{equation}%
where $\tilde{\rho}_{0}=|\mathrm{TFD}\rangle \langle \mathrm{TFD}|$ is of
dimension $d^{2}$. Said differently, $F_{t}$ equals the probability to find
the state $\tilde{\rho}_{t}$ at time $t$ in the initial state, i.e., it is the
survival probability of the TFD state. Note that under unitary evolution, $%
\Lambda _{t}[\tilde{\rho}_{0}]=U_{t}\tilde{\rho}_{0}U_{t}^{\dag }$ and Eq. (%
\ref{FidelityU}) is recovered.

For the sake of illustration, we shall consider the quantum channel
associated with energy diffusion processes that occur independently in each
of the copies. The total Hamiltonian $\tilde{H}=H\otimes \mathds{1}+%
\mathds{1}\otimes H$ is perturbed by independent real Gaussian white noise
in each copy, $H\rightarrow (1+\sqrt{\gamma }\xi _{t})H$, where $\gamma $ is
a positive real constant, and $\xi _{t}$ is the noise parameter. Performing the stochastic average, the evolution of $\tilde{\rho}_{t}$
is described by the  master equation  \cite{Xu19,delcampo19}
\begin{equation}
\dot{\tilde{\rho}}_{t}=-i\left[ \tilde{H},\tilde{\rho}_{t}\right] -\frac{%
\gamma }{2}\sum_{k}\left[ V_{k},\left[ V_{k},\tilde{\rho}%
_{t}\right] \right] ,  \label{master eq}
\end{equation}%
with the Lindblad operators $V_{1}=H\otimes \mathds{1}$ and $V_{2}=\mathds{1}%
\otimes H$.

For the initial TFD state, the exact time evolution of the density matrix is
given by
\begin{equation}
\tilde{\rho}_{t}=\sum_{m,n}\frac{e^{-\frac{\beta (E_{m}+E_{n})}{2}}}{Z(\beta
)}e^{-2it(E_{m}-E_{n})-\gamma t(E_{m}-E_{n})^{2}}|mm\rangle \!\left\langle
nn\right\vert ,
\end{equation}%
and the fidelity (\ref{Fidelity}) of the evolved mixed state reads
\begin{equation}
F_{t}=\frac{1}{Z(\beta )^{2}}\sum_{m,n}e^{-\beta
(E_{m}+E_{n})-2it(E_{m}-E_{n})-\gamma t(E_{m}-E_{n})^{2}}.  \label{Fidelityt}
\end{equation}%
From this expression it is apparent that, in the absence of degeneracies in
the energy spectrum, the asymptotic value of the fidelity is given by $%
F_{p}=Z(2\beta )/Z(\beta )^{2}$, i.e., the purity of a single-copy thermal
state at inverse temperature $\beta $. This value also corresponds to the
long-time asymptotics under unitary evolution, which can be obtained from
Eq. (\ref{FidelityU}) by coarse-graining in time \cite{Barbon14,delcampo17}.
In the infinite temperature case, the value $F_{p}=1/d$ reflects the finite Hilbert
space dimension. Thus, $F_{p}$ is insensitive to the presence of
information loss.

For arbitrary time $t$, an explicit expression of the fidelity can be
obtained using the density of states $\varrho (E)=\sum \delta (E-E_{n})$
written in the integral form, $\varrho (E)=\int d\tau e^{i\tau E}\mathrm{Tr}%
(e^{-i\tau H})/(2\pi )$. Use of the
Hubbard--Stratonovich transformation allows us to recast the fidelity (\ref%
{Fidelityt}) in terms of the analytic continuation of the partition
function  \cite{SM},
\begin{equation}
F_{t}=\frac{1}{2\sqrt{\pi \gamma t}}\int_{-\infty }^{+\infty }d\tau
e^{-\left( \frac{\tau -2t}{2\sqrt{\gamma t}}\right) ^{2}}g_{\beta }(\tau ),
\label{fid2}
\end{equation}%
as the spectral form factor is given by
\begin{equation}
g_{\beta }(\tau )\equiv \frac{|Z(\beta +i\tau )|^{2}}{Z^{2}(\beta )},
\end{equation}%
and equals the fidelity under unitary dynamics at $\tau=2t$, see Eq. (\ref{FidelityU}).
The latter is an even function of the parameter $\tau $, i.e., $g_{\beta
}(-\tau )=g_{\beta }(\tau )$. This quantity contains information about the
correlation of eigenvalues with different energies. At late times, it forms
a plateau, with a value $Z(2\beta )/Z(\beta )^{2}$ in absence of
degeneracies in the energy spectrum, that characterizes the discreteness of
the spectrum \cite{Cotler2017}.

The expression (\ref{fid2}) paves the way to a systematic study of the
interplay between quantum chaos and information loss, provided the energy spectrum of the system is known. In addition, it shows that
noise-induced decoherence is equivalent to coarse-graining in time the
spectral form factor with a specific Gaussian kernel. The quantity $\sqrt{%
\gamma t}$ determines the contribution of the spectral form factor to the
integral at any time $t$, i.e., to the fidelity. In the unitary limit, $%
\gamma =0$, the Gaussian is sharply peaked and tends to a Dirac delta around
$2t$, leading to the recovery of the spectral form factor in Eq. (\ref%
{FidelityU}).  For $\gamma >0$, information is lost. Yet, at
long times of evolution, the behavior with and without decoherence agree.
This is consistent with the fact that the long-time asymptotic plateau is
associated with a state diagonal in the energy eigenbasis. The later is the
fixed point under the nonunitary dephasing dynamics considered but it also
emerges effectively under unitary dynamics
with  coarse-graining in time. The effect of dephasing is thus crucial in the
time region $\gamma t\sim 1$. This behavior is universal in that it arises
from the open quantum dynamics considered (\ref{master eq}) and is
independent of the specific choice of the nondegenerate system Hamiltonian.

\textit{The Sachdev-Ye-Kitaev model.---}
In what follows we shall use as a test-bed the SYK model
with Hamiltonian \cite{SachdevYe93,Kitaev15}
\begin{equation}
H=\sum_{1\leq k<l<m<n\leq N}J_{klmn}\chi _{k}\chi _{l}\chi _{m}\chi _{n},
\label{SYK}
\end{equation}%
describing $N$ Majorana fermions, $\chi _{k}=\chi _{k}^{\dag }$
fulfilling the anti-commutation relation $\{\chi _{k},\chi _{l}\}=\delta
_{kl}$, with an all-to-all random quartic interaction,
and couplings
independently sampled from a Gaussian distribution $P(J_{klmn})\propto \exp %
\big(-\frac{N^{3}}{12J^{2}}J_{klmn}^{2}\big)$, centered at zero, $\overline{%
J_{klmn}}=0$, with variance $\overline{J_{klmn}^{2}}=\frac{%
3!J^{2}}{N^{3}}$.
The results that follow always include the disorder average. We set $J=1$
for convenience.
Being a test-bed for holography and quantum chaos,
 is amenable to digital quantum simulation in a variety of
platforms including trapped ions, superconducting qubits and NMR experiments
\cite{Laura17,Babbush19,Luo19}. Its study in analogue simulators has also
been proposed \cite{Danshita17,Pikulin17,Franz18,Sedrakyan20}.
Spectral properties of SYK model can be
captured by different random matrix ensembles depending on the value of $N$
\cite{Garcia-Garcia2016PRD,Cotler2017,You2017}. We consider $N=26$ in the numerics, when
spectral features are captured by the Gaussian unitary ensemble \cite%
{MethaBook,SM}. 
\begin{figure}[t]
\begin{center}
\includegraphics[width=\linewidth]{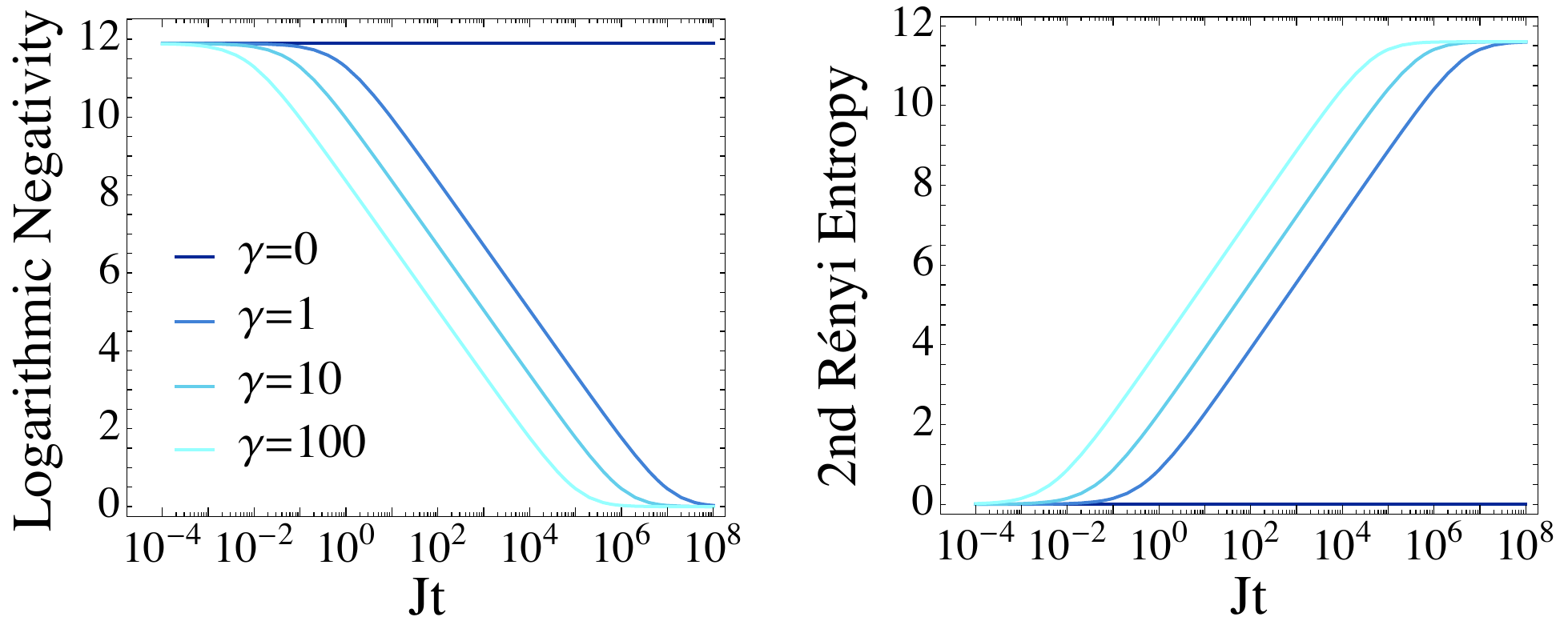}
\end{center}
\caption{\textbf{Logarithmic negativity and R\'{e}nyi entropy of the
stochastic SYK model.} A log-linear plot of the (a) logarithmic negativity
and (b) R\'{e}nyi entropy (n=2) displayed as a function of time for
different decoherence coefficients $\protect\gamma $ in the stochastic SYK
model with $N=26$ Majorana Fermions. The data is built from 100 independent
samples and $\protect\beta =1$.}
\label{FlucSYKRenyi}
\end{figure}

We first consider the role of the open dynamics on the logarithmic
negativity, an entanglement monotone defined as $E_{N}(\tilde{\rho}_{t})=\log
_{2}\Vert \tilde{\rho} _{t}^{\Gamma _{L}}\Vert _{1}$ in terms of the trace norm of
the partial transpose of the density matrix $\tilde{\rho} _{t}^{\Gamma _{L}}$, e.g.,
with respect to the left copy. Figure \ref{FlucSYKRenyi} shows that as a
function of time this quantity exhibits a monotonic decay, signaling the
loss of entanglement between the two copies of the TFD state. By contrast, $%
E_{N}(\tilde{\rho} _{t})$ remains constant under the unitary evolution with $\gamma
=0$. The loss of entanglement is
accompanied by information loss, manifested by the monotonic growth of the
second R\'{e}nyi entropy $S_{2}(\tilde{\rho} _{t})=-\log _{2}\mathrm{Tr}(\tilde{\rho}
_{t}^{2})$ that stems exclusively from the nonunitary features of the
dynamics encoded in the dissipator; see Fig. \ref{FlucSYKRenyi}. Again, this
quantity is invariant under unitary dynamics.
The relevant time scale governing their evolution is the decoherence time
that can be extracted from the short-time asymptotics of the purity, $P_{t}=%
\mathrm{Tr}(\tilde{\rho} _{t}^{2})$, which is invariant under unitary
dynamics. The time scale $\tau _{D}$ governs
its short-time asymptotics according to $P_{t}=1-t/\tau _{D}+\mathcal{O}%
(t^{2})$ \cite{Shimizu02,Beau17,Xu19}. The same behavior, replacing $\tau
_{D}$ by $2\tau _{D}$, rules the early decay of the fidelity due to the
nonunitary character of the evolution, i.e., $F_{t}=1-t/({2\tau _{D}})+%
\mathcal{O}(t^{2})$ \cite{Lidar98,Chenu17}. The time scale in which the TFD
becomes diagonal in the energy eigenbasis is set by \cite{Xu19}
\begin{equation}
\frac{1}{\tau _{D}}=4\gamma \frac{\mathrm{d}^{2}}{\mathrm{d}\beta ^{2}}\ln %
\left[ Z(\beta )\right] .  \label{taud}
\end{equation}%
Using a Gaussian approximation for the  density of states $\varrho (E)$ at large $N$
\cite{Garcia-Garcia2016PRD,Garcia-Garcia2017PRD,Liu2017}, one finds $\tau _{D}=1/(\gamma N)$ \cite{SM}.
The monotonic decay of $E_{N}(\tilde{\rho} _{t})$ and $S_{2}(\tilde{\rho} _{t})$ prevents an
investigation of the competition between quantum chaos signatures and
information loss. To this end, we rely on the use of the fidelity $F_{t}$.

Under unitary evolution, the fidelity of the TFD state in the
SYK model exhibits the typical features of chaotic quantum systems \cite%
{Cotler2017,Dyer2017,delcampo17}, namely a decay, dip, ramp and plateau. The
existence of a plateau is a consequence of the Hilbert space finite dimension,
in which the energy spectrum is discrete. It is absent in systems with a
continuous spectrum, where the decay is uninterrupted and continues to a
vanishing value \cite{Fock47,Beau17}. As a function of the time of
evolution, the behavior of the fidelity is first dominated by (i) the
density of states and decays from unity until it reaches a minimum value at
a dip occurring at the 
dip time $t_{d}\sim \sqrt{d}$, where $d$ is the Hilbert
space dimension of a single copy of the system. In the SYK, we estimate the
dip time as
\begin{equation}
t_{d}\sim \left( \frac{\sqrt{\pi }\exp \left( -N\beta ^{2}/4\right) }{%
c_{N}^{3/2}\sqrt{2N}}\right) ^{1/4}\sqrt{d},
\end{equation}%
with $c_{N}$ a constant \cite{SM}. The subsequent time evolution is
dominated by correlations in the eigenvalue spacing and leads to (ii) a ramp
that eventually saturates in (iii) a plateau with value $F_{p}=Z(2\beta
)/Z(\beta )^{2}$ onset at the Heisenberg or plateau time $t_{p}\sim d$. Specifically, for the SYK model we find \cite{SM}
\begin{equation}
t_{p}\simeq \alpha \sqrt{\frac{2\pi }{N}}d,
\end{equation}%
with $\alpha =2-\delta _{4,N\,\text{mod}\,8}$. This late stage is
characterized by fluctuations around the plateau value, sometimes refer to as
quantum noise in this context \cite{Barbon14} to be distinguished from the
kind of quantum noise that gives rise to decoherence \cite%
{GardinerZollerBook}. %
\begin{figure}[tbp]
\begin{center}
\includegraphics[width=\linewidth]{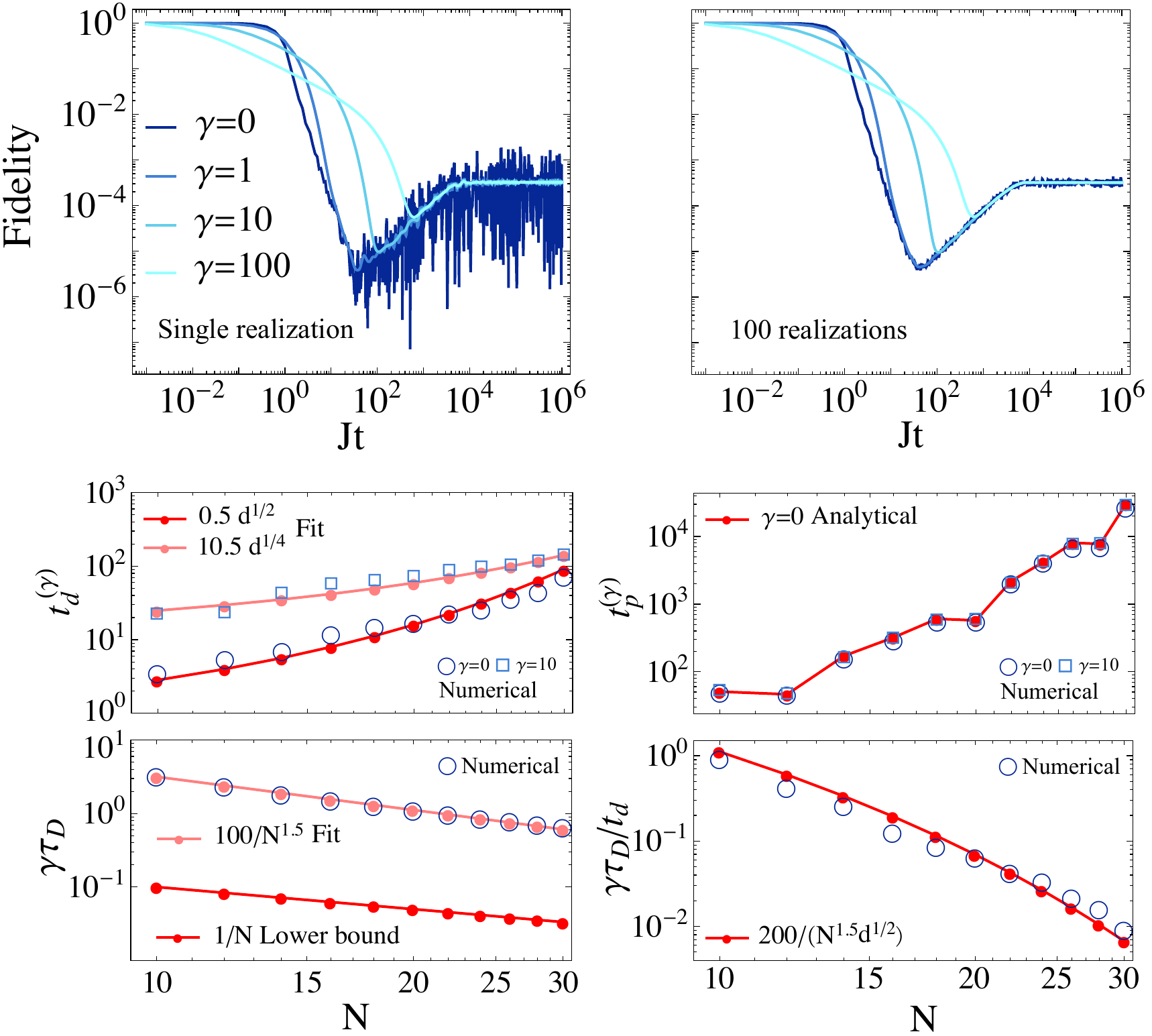}
\end{center}
\caption{\textbf{Fidelity of the stochastic SYK model.} Top: A log-log plot of fidelity with different decoherence coefficients $%
\protect\gamma $ in the SSYK model of $N=26$ Majorana Fermions. The data was
taken by single and 100 independent samples and $\protect\beta =1$. Bottom: The dip
time $t_{d}^{(\protect\gamma )}$, the plateau time $t_{p}^{(\protect%
\gamma )}$, the decoherence time $\protect\gamma \protect\tau _{D}$, and $\protect\gamma \protect\tau _{D}$/$t_{d}$ are shown as a function of $N$.}
\label{FlucSYK G}
\end{figure}
%
The characteristic times $\tau_D$, $t_d$ and $t_p$ govern the competition between decoherence and quantum chaos.

Figure \ref{FlucSYK G} shows the evolution of the fidelity for
a finite-temperature TFD in a single realization and the disorder average over $J_{klmn}$. As the dephasing
strength $\gamma $ is increased, the features of the fidelity $F_{t}$
manifested in the unitary case are gradually washed out. Prominently, for
large dephasing strengths, $\tau _{D}\ll t_{p}$, the existence of the
dip and ramp are completely suppressed and the fidelity decays monotonically
from unit value towards the asymptotic one $F_{p}$. Between
these extremes, the features that are more sensitive to
information loss are those associated with quantum noise, e.g., the
dynamical phase accumulated by the total Hamiltonian.

Under unitary
dynamics, these fluctuations are exhibited around the dip and at long times:
The decay towards the dip is typically characterized by a power-law given that the density of states is bounded from below, i.e., the existence of a ground state \cite%
{delcampo17} (while this effect can be removed by smooth spectral filtering \cite{Lev}). 
 As a precursor of the dip, an oscillatory behavior is often
present that can be understood as the interference of the power-law
contribution and the ramp contribution stemming from correlations in the
level spacing distribution.
 Whenever $\tau _{D}\leq t_{d}$,
information loss leads to the suppression of these fluctuations. Regarding
the presence of quantum noise at long times, whenever $\tau _{D}<t_{p}$,
equation (\ref{Fidelityt}) shows that information loss associated with
decoherence suppresses the fluctuations around the plateau value $F_{p}$.
 Importantly, the suppression of quantum noise fluctuations is
already manifest at the level of a single realization of the SYK
Hamiltonian, without averaging over $J_{klmn}$ or ensembles of system
Hamiltonians. As shown in \cite{SM} the behavior of the SYK models is
 in qualitative agreement with that of random-matrix ensembles. While under
unitary dynamics this correspondence is only established at long times, its
onset is facilitated by the presence of information loss. The decoherence
time $\tau _{D}$ scales with $1/\gamma $ and the inverse energy variance. It
thus decreases with temperature and the system size, as shown  for the SYK in Fig \ref{FlucSYK G}. In the presence of information loss, the dip not
only becomes shallower, but it shifts to later times; see Fig \ref{FlucSYK G}. For moderate values of the dephasing strength $%
\gamma $ the subsequent ramp is essentially unaffected with respect to the
unitary dynamics, beyond the
suppression of quantum noise. The time scale $t_{p}$
in which the plateau appears remains essentially constant. Thus, the ramp
and plateau are shared by isolated and decohering systems exhibiting information loss.

\textit{\ Discussion and summary.---} An experimental test of the interplay
between quantum chaos and decoherence can be envisioned given advances in the quantum simulation of open systems by digital methods \cite{Barreiro2011} and tailored noise.
It could be probed  via the quantum simulation of the SYK Hamiltonian \cite%
{Danshita17,Laura17,Babbush19,Luo19} but it is generally expected in an arbitrary  quantum chaotic system. While the preparation of the TFD
state is being pursued \cite{Cottrell19,WuHsieh19,Zhu19,Miceli19} this
requirement  can  be relaxed for the study of some observables, such as the fidelity of a
TFD state, as its expectation value  can be related to that of a coherent Gibbs state $|\psi _{\beta }\rangle
=\sum_{n}e^{-\beta E_{n}/2}|n\rangle /\sqrt{Z(\beta )}$ involving a single
copy of the system. Indeed, under unitary time evolution $F_{t}=|\langle
\psi _{\beta }|\exp (-itH )\psi _{\beta }\rangle |^{2}=|Z(\beta
+it)/Z(\beta )|^{2}$ that can be measured by single-qubit interferometry \cite{Peng15}. Its general time-evolution can be described by a
quantum channel $\rho _{t}=\Lambda _{t}[|\psi _{\beta }\rangle \langle \psi
_{\beta }|]$ and the fidelity between the initial state and its evolved form
is analogously given by $F_{t}=\langle \psi _{\beta }|\rho _{t}|\psi _{\beta
}\rangle $. The measurement of the later can be simplified using quantum
algorithms for the estimation of state overlaps \cite{Cincio2018,Cerezo2020}.

In summary, the ubiquity of noise sources gives rise to a competition
between the signatures of quantum chaotic dynamics expected for a many-body
system in isolation and the presence of information loss resulting from
decoherence. Such competition can be quantified by the fidelity between a
thermofield state at a given time and its subsequent time evolution. For a
quantum chaotic system in isolation this quantity equals the spectral form
factor showing a dip-ramp-plateau structure which is suppressed by the loss
of information induced by decoherence. The interplay between information loss
and scrambling  in open quantum complex systems
 should find broad applications in quantum computation,  simulation and  machine learning in the presence of noise.

\textit{Acknowledgements.---} It is a pleasure to acknowledge discussions
with Julian Sonner, Tadashi Takayanagi, and Jacobus Verbaarschot. TP acknowledges support by ERC Advanced grant 694544 OMNES and the program P1-0402 of Slovenian Research Agency. This work is further supported by ID2019-109007GA-I00.
%

\bibliography{references_SSYK}

\newpage \appendix

\clearpage

\widetext

\setcounter{equation}{0} \setcounter{figure}{0} \setcounter{table}{0}
\makeatletter
\renewcommand{\theequation}{S\arabic{equation}} \renewcommand{\thefigure}{S%
\arabic{figure}} \renewcommand{\bibnumfmt}[1]{[#1]} \renewcommand{%
\citenumfont}[1]{#1}

\section{I. Fidelity in terms of spectral form factor}

We start with the expression of the fidelity in Eq. (\ref%
{Fidelityt}), and use the Hubbard--Stratonovich transformation
\begin{equation}
e^{-\gamma t(E_{m}-E_{n})^{2}}=\frac{1}{2\sqrt{\pi \gamma t}}\int_{-\infty
}^{+\infty }dye^{-\frac{y^{2}}{4\gamma t}}e^{-iy(E_{m}-E_{n})},  \label{HS}
\end{equation}%
to obtain a universal expression related to the normalized spectral form
factor. The fidelity with Eq. (\ref{HS}) is of the form
\begin{eqnarray}
F_{t} &=&\frac{1}{2\sqrt{\pi \gamma t}}\int_{-\infty }^{+\infty }dye^{-\frac{%
y^{2}}{4\gamma t}}\frac{1}{Z^{2}(\beta )}\sum_{m,n}e^{-(\beta
+2it+iy)E_{m}}e^{-(\beta -2it-iy)E_{n}}  \notag \\
&=&\frac{1}{2\sqrt{\pi \gamma t}}\int_{-\infty }^{+\infty }dye^{-\frac{y^{2}%
}{4\gamma t}}\frac{\left\vert Z\left( \beta +2it+iy\right) \right\vert ^{2}}{%
Z^{2}(\beta )}.  \label{Ft1}
\end{eqnarray}%
By changing the variable $\tau =y+2t$, Eq. (\ref{Ft1}) can be further
simplified to
\begin{equation}
F_{t}=\frac{1}{2\sqrt{\pi \gamma t}}\int_{-\infty }^{+\infty }d\tau
e^{-\left( \frac{\tau -2t}{2\sqrt{\gamma t}}\right) ^{2}}g_{\beta }(\tau ),%
\text{ }  \label{FtSF1}
\end{equation}%
with the spectral form factor%
\begin{equation}
g_{\beta }(\tau )\equiv \frac{\left\vert Z\left( \beta +i\tau \right)
\right\vert ^{2}}{Z^{2}(\beta )},
\end{equation}%
which is the expression given in Eq. (\ref{fid2}) of the main text.

\section{II. Logarithmic negativity and the second R\'{e}nyi entropy}

The logarithmic negativity is defined as
\begin{equation}
E_{N}(\tilde{\rho}_{t})=\log _{2}\left\Vert\tilde{\rho}_{t}^{\Gamma _{L}}\right\Vert _{1},
\label{LN}
\end{equation}%
in terms of the trace norm of the partial transpose of the density matrix,
while the second R\'{e}nyi entropy is

\begin{equation}
S_{2}(\tilde{\rho}_{t})=-\log _{2} \mathrm{tr}\tilde{\rho}_{t}^{2}.\quad  \label{RE}
\end{equation}

If the initial state is the thermal field double state, i.e.,
\begin{equation}
\tilde{\rho}_{0}=|\mathrm{TFD}\rangle \langle \mathrm{TFD}|=\frac{1}{Z(\beta )}%
\sum_{k,\ell }e^{-\frac{\beta }{2}(E_{k}+E_{\ell })}|k\rangle |k\rangle
\langle \ell |\langle \ell |,
\end{equation}%
the above Eqs. (\ref{LN}) and (\ref{RE}) can be written as
\begin{equation}
E_{N}(\tilde{\rho}_{t})=\log _{2}\left[ \frac{1}{Z(\beta )}\sum_{k\ell }e^{-\beta
(E_{k}+E_{\ell })/2-\gamma t(E_{k}-E_{\ell })^{2}}\right] ,
\end{equation}%
and
\begin{equation}
S_{2}(\tilde{\rho}_{t})=-\log _{2} \left(P_{t}\right) \text{ with }P_{t}=\frac{1}{%
Z(\beta )^{2}}\sum_{k,\ell }e^{-\beta (E_{k}+E_{\ell })-2\gamma
t(E_{k}-E_{\ell })^{2}}.
\end{equation}%
Then we have
\begin{equation}
E_{N}(\tilde{\rho} _{t}(2\beta ,2\gamma ))+S_{2}(\tilde{\rho} _{t}(\beta ,\gamma ))=\log _{2}%
\frac{Z(\beta )^{2}}{Z(2\beta )}.
\end{equation}%
Thus, the growth of the  logarithmic negativity implies the decay of the second R\'{e}%
nyi entropy, and viceversa, i.e., $\dot{E}_{N}(\tilde{\rho} _{t}(2\beta ,2\gamma
))=-\dot{S}_{2}(\tilde{\rho} _{t}(\beta ,\gamma ))$.

\section{III. Fidelity in terms of density of states and form factor}

In what follows, we make explicit the connection between the fidelity, the density
of states and the spectral form factor. The density of states is
defined as
\begin{equation*}
\varrho (E)=\sum_{n}N_{n}\delta (E-E_{n}),
\end{equation*}%
where $N_{n}$ denotes the degeneracy of the energy level $E_{n}$.

The thermal state of a single copy can be written as
\begin{equation*}
\rho _{\rm th }=\frac{1}{Z(\beta )}\int dE\sigma (E)e^{-\beta E}|E\rangle
\langle E|.
\end{equation*}%
Its purification is given by the thermofield double state
\begin{equation*}
|\Psi \rangle =\frac{1}{\sqrt{Z(\beta )}}\int dE\sqrt{\varrho (E)}e^{-\beta
E/2}|E,E\rangle .
\end{equation*}%
The initial density matrix associated with the thermofield double state can
then be written using a basis of continuous energy eigenstates as
\begin{equation*}
\tilde{\rho} _{0}(E,E^{\prime })=\int dEdE^{\prime }\frac{e^{-\beta (E+E^{\prime
})/2}}{Z(\beta )}\sqrt{\varrho (E)\varrho (E^{\prime })}|E,E\rangle \langle
E^{\prime },E^{\prime }|.
\end{equation*}%
In turn, the time-evolved density matrix reads
\begin{eqnarray}
\tilde{\rho} _{t}(E,E^{\prime }) &=&\int dEdE^{\prime }\frac{e^{-\beta (E+E^{\prime
})/2}}{Z(\beta )}\sqrt{\varrho (E)\varrho (E^{\prime })}e^{-2it(E-E^{\prime
}) }e^{-\gamma t(E-E)^{\prime 2}/2}|E,E\rangle \langle E^{\prime
},E^{\prime }| \\
&=&\sqrt{\frac{1}{4\pi \gamma t}}\int dEdE^{\prime }\int_{-\infty }^{\infty
}dy e^{-\frac{y^{2}}{4\gamma t}}\frac{e^{-\beta (E+E^{\prime })/2}}{Z(\beta )}%
\sqrt{\varrho (E)\varrho (E^{\prime })}e^{-2it(E-E^{\prime })
}e^{-iy(E-E^{\prime })}|E,E\rangle \langle E^{\prime },E^{\prime }|
\end{eqnarray}%
and the fidelity becomes
\begin{equation*}
F_{t}=\sqrt{\frac{1}{4\pi \gamma t}}\int_{-\infty }^{\infty }dy e^{-\frac{y^{2}%
}{4\gamma t}}\frac{1}{Z(\beta )^{2}}\left\vert \int dE\sigma (E)e^{-(\beta
-2it +iy)E}\right\vert ^{2}.
\end{equation*}

The evaluation of such expression generally requires the use of numerical
methods due to the lack of techniques to evaluate the average of the
quotient of partition functions, each of which involving the Hamiltonian
over which the average is performed. Under the annealed approximation, this
average is approximated by the quotient of the averages. This approximation
is generally valid at high temperature and fails at low temperature. Using
it, the average fidelity reads
\begin{equation*}
\langle F_{t}\rangle =\sqrt{\frac{1}{4\pi \gamma t}}\int_{-\infty }^{\infty
}dy e^{-\frac{y^{2}}{4\gamma t}}\frac{1}{\langle Z(\beta )^{2}\rangle }\int
dEdE^\prime e^{ -\nu (y)E-\bar{\nu}(y)E^{\prime }}\langle \varrho
^{(2)}(E,E^{\prime })\rangle ,
\end{equation*}%
where
\begin{equation*}
\nu (y)=\beta -2it +iy,\quad \bar{\nu}(y)=\beta +2it -iy.
\end{equation*}%
The two-level correlation function $\langle \varrho ^{(2)}(E,E^{\prime
})\rangle =\langle \varrho (E)\varrho (E^{\prime })\rangle $ can be
expressed in terms of the connected two-level correlation function $\langle
\varrho _{c}^{(2)}(E,E^{\prime })\rangle $
\begin{equation*}
\langle \varrho _{c}^{(2)}(E,E^{\prime })=\langle \varrho (E,E^{\prime })\rangle
-\langle \varrho (E)\rangle \langle \varrho (E^{\prime })\rangle .
\end{equation*}

Assuming no degeneracies
\begin{equation}
F_{t}=\frac{G(\beta )}{Z(\beta )^{2}}+\frac{1}{Z(\beta )^{2}}\sum_{k\neq
\ell }e^{-\beta (E_{k}+E_{\ell })+i2t(E_{k}-E_{\ell })-\gamma
t(E_{k}-E_{\ell })^{2}}.
\end{equation}%
where
\begin{equation*}
G(\beta )=\sum_{n}N_{n}^{2}e^{-2\beta E_{n}}=\int dE\langle \varrho
^{(2)}(E,E)\rangle e^{-2\beta E},
\end{equation*}%
reduces to $Z(2\beta )$ in the absence of degeneracies expected for chaotic
systems, i.e., when $N_{n}=1$.

\section{IV. Ensemble average of the fidelity}

The ensemble average over the fidelity of Eq. (\ref{fid2}) in the main text
can be written as
\begin{equation}
\left\langle F_{t}\right\rangle =\frac{1}{2\sqrt{\pi \gamma t}}\int_{-\infty
}^{+\infty }d\tau e^{-\left( \frac{\tau -2t}{2\sqrt{\gamma t}}\right)
^{2}}\left\langle g_{\beta }(\tau )\right\rangle ,\text{ }
\end{equation}%
where the averaged spectral form factor in terms of annealing approximation
is given by
\begin{equation}
\left\langle g_{\beta }(\tau )\right\rangle \doteq \frac{\left\langle
\left\vert Z\left( \beta +i\tau \right) \right\vert ^{2}\right\rangle }{%
\left\langle Z(\beta )\right\rangle ^{2}}.  \label{averaged-g}
\end{equation}%
Note that the annealing approximation is valid at high temperature (also see
discussions in the end of Sec. V-A).

The density of states is defined as
\begin{equation*}
\varrho (E)=\sum_{n}N_{n}\delta (E-E_{n}),
\end{equation*}%
where $N_{n}$ denotes the degeneracy of the energy level $E_{n}$. Then the
denominator and numerator of Eq. (\ref{averaged-g}) can be written as%
\begin{equation}
\left\langle Z(\beta )\right\rangle =\int dE\langle \varrho (E)\rangle
e^{-\beta E},  \label{averaged-z}
\end{equation}%
and
\begin{equation}
\left\langle \left\vert Z\left( \beta +i\tau \right) \right\vert
^{2}\right\rangle =\int dE\langle \varrho (E)^{2}\rangle e^{-2\beta E}+\int
dEdE^{\prime }\langle \varrho (E)\varrho (E^{\prime })\rangle e^{-(\beta
+i\tau )E}e^{-(\beta -i\tau )E^{\prime }},  \label{averaged-zz}
\end{equation}%
where the two-point correlation function can be expressed as
\begin{equation}
\langle \varrho (E)\varrho (E^{\prime })\rangle =\langle \varrho
_{c}^{(2)}(E,E^{\prime })\rangle +\left\langle \varrho (E)\right\rangle \langle
\varrho (E^{\prime })\rangle ,
\end{equation}%
in terms of the connected two-point correlation function $\varrho
_{c}(E,E^{\prime })$.

In the following, we consider two examples, one is the GUE, and the other is
the SYK model.

\subsection{A. GUE-averaged Fidelity}

For GUE ensembles, there are no degeneracy of the energy levels, so Eq. (\ref%
{averaged-zz}) can be further written as%
\begin{equation}
\left\langle \left\vert Z\left( \beta +i\tau \right) \right\vert
^{2}\right\rangle _{\mathrm{GUE}}=\left\langle Z(2\beta )\right\rangle _{%
\mathrm{GUE}}+\left\vert \left\langle Z(\beta +i\tau )\right\rangle _{%
\mathrm{GUE}}\right\vert ^{2}+\left\langle g_{\beta }^{c}(\tau
)\right\rangle _{\text{\textrm{GUE}}},  \label{GUEzz}
\end{equation}%
where%
\begin{equation}
\left\langle g_{\beta }^{c}(\tau )\right\rangle _{\text{\textrm{GUE}}}=\int
dEdE^{\prime }\langle \varrho _{c}^{(2)}(E,E^{\prime })\rangle _{\mathrm{GUE}%
}e^{-(\beta +i\tau )E}e^{-(\beta -i\tau )E^{\prime }}.  \label{gc}
\end{equation}%
\ \ \ \ \ \ \

The joint probability density of $H\in $\textrm{GUE} is proportional to $%
\exp (-\frac{1}{2\sigma ^{2}}\mathrm{tr}H^{2})$, where $\sigma $ is the
standard deviation of the random (off-diagonal) matrix elements of $H$. Note that in Ref. \cite%
{MethaBook}, $\sigma =1/\sqrt{2}$, and in Ref. \cite{Cotler2017}, $\sigma =1/%
\sqrt{d}$. To calculate Eq. (\ref{GUEzz}), we have to know the spectral
density and the two-point correlation function. The eigenvalue density
averaged over \textrm{GUE} is given by
\begin{equation}
\left\langle \varrho (E)\right\rangle _{\mathrm{GUE}}=\frac{1}{\sqrt{2}%
\sigma }K_{d}\left( \tilde{E},\tilde{E}\right) ,\text{ and }\tilde{E}:=\frac{%
E}{\sqrt{2}\sigma }\text{,}  \label{GUEz}
\end{equation}%
with the kernel $K_{d}(x,y)$ defined by

\begin{equation}
K_{d}(x,y)=\sum_{l=0}^{d-1}\phi _{l}(x)\phi _{l}(y)\text{, and }\phi
_{l}(x):=\frac{e^{-\frac{x^{2}}{2}}\mathrm{H}_{l}(x)}{\sqrt{\sqrt{\pi }%
2^{l}l!}},  \label{kernel}
\end{equation}%
where $\mathrm{H}_{l}(x)$ are the Hermite polynomials. Furthermore, the
two-point correlation function averaged over \textrm{GUE} takes the form%
\begin{equation}
\langle \varrho (E,E^{\prime })\rangle _{\mathrm{GUE}}=\frac{1}{2\sigma ^{2}}%
\det \left[ \left(
\begin{array}{ll}
K_{d}(\tilde{E},\tilde{E}) & K_{d}(\tilde{E},\tilde{E}^{\prime }) \\
K_{d}(\tilde{E}^{\prime },\tilde{E}) & K_{d}(\tilde{E}^{\prime },\tilde{E}%
^{\prime })%
\end{array}%
\right) \right] ,
\end{equation}%
and thus the connected two-level correlation function averaged over GUE
reads
\begin{equation}
\langle \varrho _{c}^{(2)}(E,E^{\prime })\rangle _{\mathrm{GUE}}=-\frac{1}{2\sigma
^{2}}\left( K_{d}(\tilde{E},\tilde{E}^{\prime })\right) ^{2}.  \label{GUEc}
\end{equation}

According to the orthogonality of Hermite polynomials%
\begin{equation}
\int dxe^{-(x+a)^{2}}\mathrm{H}_{k}(x)\mathrm{H}_{l}(x)\text{=}\sqrt{\pi }%
2^{p}p!(-2a)^{|k-l|}\mathrm{L}_{p}^{(|k-l|)}(-2a^{2})\text{, }  \label{HH}
\end{equation}%
where $\mathrm{L}_{n}^{(\alpha )}(\cdot )$ are the associated Laguerre
polynomials, and $p$:=min\{$k,l$\}, the first two items of Eq. (\ref{GUEzz})
and the denominator of Eq. (\ref{averaged-g}) is expressed as%
\begin{equation}
\left\langle Z(x)\right\rangle _{\mathrm{GUE}}=e^{\frac{\sigma ^{2}x^{2}}{2}}%
\mathrm{L}_{d-1}^{(1)}\left( -\sigma ^{2}x^{2}\right) .  \label{gz}
\end{equation}%
By Eqs. (\ref{GUEc}) and (\ref{HH}), the third term of Eq. (\ref{GUEzz}) is
\begin{equation}
\left\langle g_{\beta }^{c}(\tau )\right\rangle _{\text{\textrm{GUE}}%
}=-e^{\sigma ^{2}(\beta ^{2}-\tau ^{2})}\sum_{k,l=0}^{d-1}\frac{p!}{q!}%
\left( \sigma ^{2}(\beta ^{2}+\tau ^{2})\right) ^{|k-l|}\left\vert \mathrm{L}%
_{p}^{(|k-l|)}\left( -\sigma ^{2}(\beta +i\tau )^{2}\right) \right\vert ^{2},
\label{gc2}
\end{equation}%
where $q:=$max\{$k,l$\}.

With Eqs. (\ref{gz}) and (\ref{gc2}), the spectral form factor is finally
obtained%
\begin{equation}
\left\langle g_{\beta }(\tau )\right\rangle _{\text{\textrm{GUE}}}\doteq
\frac{e^{\sigma ^{2}\beta ^{2}}\mathrm{L}_{d-1}^{(1)}\left( -4\sigma
^{2}\beta ^{2}\right) +e^{-\sigma ^{2}\tau ^{2}}\left[ \left\vert \mathrm{L}%
_{d-1}^{(1)}\left( -\sigma ^{2}\beta _{\tau }^{2}\right) \right\vert
^{2}-\sum_{k,l=0}^{d-1}\frac{p!}{q!}\left( \sigma ^{2}\left\vert \beta
_{\tau }\right\vert ^{2}\right) ^{|k-l|}\left\vert \mathrm{L}%
_{p}^{(|k-l|)}\left( -\sigma ^{2}\beta _{\tau }^{2}\right) \right\vert ^{2}%
\right] }{\mathrm{L}_{d-1}^{(1)}\left( -\sigma ^{2}\beta ^{2}\right) ^{2}},
\label{sff-finite d}
\end{equation}%
with $\beta _{\tau }:=\beta +i\tau $ for short. Note again, one should
replace $\tau $ with $2t$ when directly analyzing the spectral form factor.

To have a rough estimation of the dip and plateau time, we will consider an
approximated connected two-level correlation function of Eq. (\ref{GUEc})
when the dimension of \textrm{GUE} is large, i.e.,
\begin{equation}
\langle \varrho _{c}^{(2)}(E,E^{\prime })\rangle _{\mathrm{GUE}}\simeq -\frac{1}{%
\pi ^{2}}\left( \frac{\sin ((E-E^{\prime })\sqrt{d}/\sigma )}{E-E^{\prime }}%
\right) ^{2}.
\end{equation}%
By defining new variables $r=E-E^{\prime }$ and $\omega =(E+E^{\prime })/2$,
Eq. (\ref{gc}) is given by%
\begin{equation}
\left\langle g_{\beta }^{c}(t)\right\rangle _{\text{\textrm{GUE}}}\simeq -%
\frac{1}{\pi ^{2}}\int d\omega e^{-2\beta \omega }\int_{-\infty }^{\infty
}dr\left( \frac{\sin (r\sqrt{d}/\sigma )}{r}\right) ^{2}e^{-2itr},
\label{gc3}
\end{equation}%
where we have replaced $\tau $ with $2t$. The first integration is
divergent. For estimation, the integration is replaced with
\begin{equation}
\int d\omega \rightarrow \int_{-\omega _{0}}^{\omega _{0}}d\omega .
\end{equation}%
Since the spectral density can be approximated by Wigner's semicircle in
large $d$ limit, i.e.,%
\begin{equation}
\left\langle \varrho (E)\right\rangle _{\mathrm{GUE}}=\frac{\sqrt{d}}{\sigma
\pi }\sqrt{1-\left( \frac{E}{2\sigma \sqrt{d}}\right) ^{2}},\text{ and }%
|E|\leq 2\sigma \sqrt{d},  \label{semiC}
\end{equation}%
thus $\left\langle \varrho (0)\right\rangle _{\mathrm{GUE}}=\sqrt{d}/(\sigma
\pi )$. According to the normalization of $\left\langle \varrho
(E)\right\rangle _{\mathrm{GUE}}$, we have $2\omega _{0}\left\langle \varrho
(0)\right\rangle _{\mathrm{GUE}}\simeq d$, and
\begin{equation}
\omega _{0}\simeq \frac{\sigma \pi \sqrt{d}}{2},
\end{equation}%
with which, the first integration reads
\begin{equation}
\int d\omega e^{-2\beta \omega }\simeq \frac{\sinh \left( \sqrt{d}\pi \beta
\sigma \right) }{\beta }.
\end{equation}%
The second integration in Eq. (\ref{gc3}) is the Fourier transform
\begin{equation}
\int_{-\infty }^{\infty }dr\left( \frac{\sin (r\sqrt{d}/\sigma )}{r}\right)
^{2}e^{-2itr}=\left\{
\begin{array}{ll}
\pi (\sqrt{d}/\sigma -t), & t\leq \sqrt{d}/\sigma , \\
0, & t>\sqrt{d}/\sigma .%
\end{array}%
\right.
\end{equation}%
Equation (\ref{gc3}) finally takes the form%
\begin{equation}
\left\langle g_{\beta }^{c}(t)\right\rangle _{\text{\textrm{GUE}}}\simeq
\left\{
\begin{array}{ll}
-\frac{\sinh \left( \sqrt{d}\pi \beta \sigma \right) }{\pi \beta }(\sqrt{d}%
/\sigma -t), & t\leq \sqrt{d}/\sigma , \\
0, & t>\sqrt{d}/\sigma .%
\end{array}%
\right.  \label{gc4}
\end{equation}%
According to the above equation, it is easy to observe that the plateau time
is
\begin{equation}
t_{p}=\sqrt{d}/\sigma .
\end{equation}

\begin{figure*}[t]
\begin{center}
\includegraphics[width=0.95\linewidth]{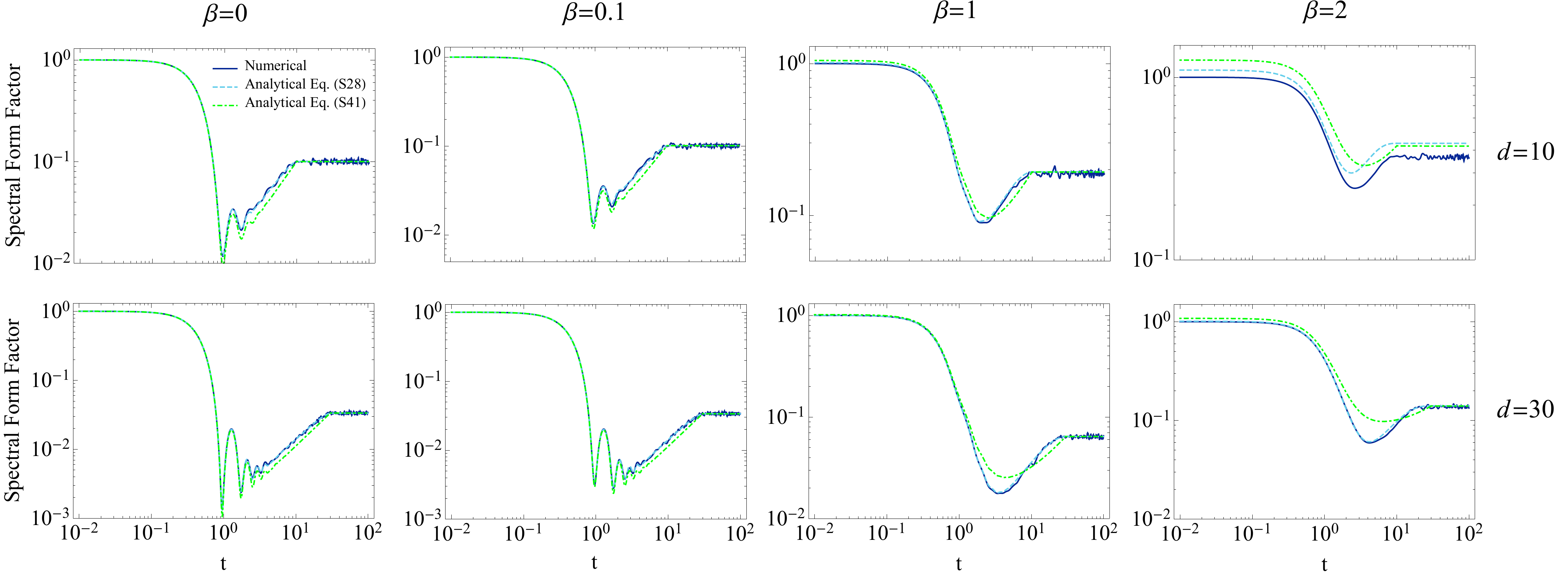}
\end{center}
\caption{\textbf{Spectral form factor of GUE.} Analytical Eqs. (\protect\ref%
{sff-finite d}) and (\protect\ref{sff-infinite d}) are in comparison with
numerical calculations for $\protect\beta =0,$ $0.1$, $1$, $2$, and $d=10$, $%
30$. The standard deviation of the random variables of $H$ is selected
as $\protect\sigma =1/\protect\sqrt{d}$. The numerical calculations use 1000 independent realizations.}
\label{SFF_GUE}
\end{figure*}

With Wigner's semicircle law of Eq. (\ref{semiC}), the partition function
averaged over the GUE ensembles is approximated by

\begin{equation}
\left\langle Z(x)\right\rangle _{\mathrm{GUE}}=\frac{\sqrt{d}\mathrm{I}%
_{1}(2\sigma \sqrt{d}x)}{\sigma x},  \label{gc4.5}
\end{equation}%
where $\mathrm{I}_{n}(\cdot )$ is the modified Bessel function of first kind
and order $n$. When $\beta \ll 1$ and $t$ is large, the asymptotic expansion
of the second part of Eq. (\ref{GUEzz}) reads%
\begin{equation}
\left\vert \left\langle Z(\beta +i2t)\right\rangle _{\mathrm{GUE}%
}\right\vert ^{2}\simeq \frac{\sqrt{d}(1-\sin (8\sigma \sqrt{d}t))}{16\pi
t^{3}\sigma ^{3}}.  \label{gc5}
\end{equation}%
By equating Eq. (\ref{gc5}) and Eq. (\ref{gc4}), the dip time can be
estimated as%
\begin{equation}
t_{d}\simeq \frac{1}{2}\left( \frac{\sqrt{d}\beta }{\sigma ^{3}\sinh \left(
\sqrt{d}\pi \beta \sigma \right) }\right) ^{\frac{1}{4}}\simeq \frac{1}{2\pi
^{1/4}\sigma }-\frac{d\pi ^{7/4}\sigma }{48}\beta ^{2}+\mathcal{O}(\beta
^{3}).
\end{equation}

The spectral form factor in the large $d$ limit is obtained
\begin{equation}
\left\langle g_{\beta }(t)\right\rangle _{\text{\textrm{GUE}}}\simeq \frac{%
\frac{\sqrt{d}\mathrm{I}_{1}(4\sigma \sqrt{d}\beta )}{2\sigma \beta }+\frac{%
\sqrt{d}(1-\sin (8\sigma \sqrt{d}t))}{16\pi t^{3}\sigma ^{3}}+\text{Eq}.(\ref%
{gc4})}{\left( \frac{\sqrt{d}\mathrm{I}_{1}(2\sigma \sqrt{d}\beta )}{\sigma
\beta }\right) ^{2}}.  \label{sff-infinite d}
\end{equation}%
In Fig. (\ref{SFF_GUE}), the spectral form factor with Eqs. (\ref{sff-finite
d}), and (\ref{sff-infinite d}) are compared with the numerical
calculations. Note that when the dimension $d$ increases, the valid domain
of $\beta $ by annealing approximation becomes larger.

\subsection{B. Spectral form factor of the SYK model}

\begin{figure*}[t]
\begin{center}
\includegraphics[width=0.85\linewidth]{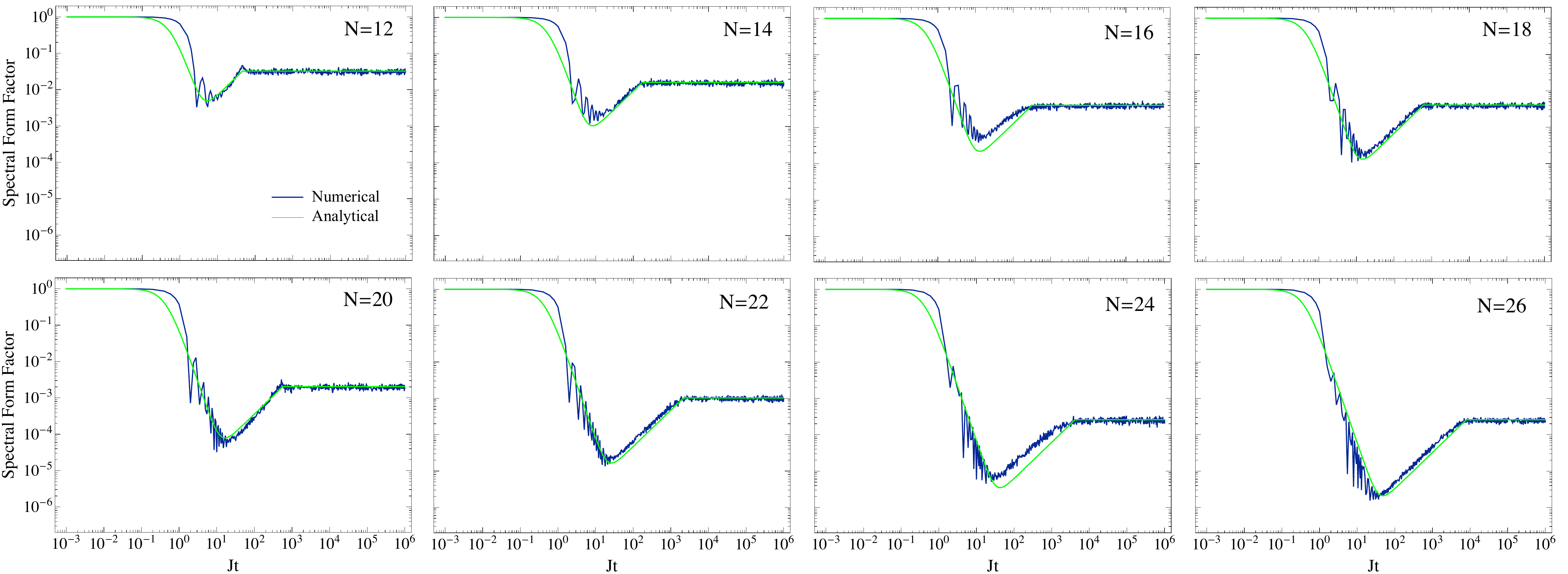}
\end{center}
\caption{\textbf{Spectral form factor of SYK model.} Analytical Eq. (\protect
\ref{gSYK}) is  commpared with numerical calculations ($\protect\beta =0.1
$). }
\label{SFF_SYK}
\end{figure*}

For SYK model with $N$ mod $8\neq 0$, the energy spectrum has a uniformly
double degeneracy ($N_{n}=2$). Equation (\ref{averaged-zz}) can be written as%
\begin{equation}
\left\langle \left\vert Z\left( \beta +i2t\right) \right\vert
^{2}\right\rangle _{\mathrm{SYK}}=2\left\langle Z(2\beta )\right\rangle _{%
\mathrm{SYK}}+\left\vert \left\langle Z(\beta +i2t)\right\rangle _{\mathrm{%
SYK}}\right\vert ^{2}+\left\langle g_{\beta }^{c}(t)\right\rangle _{\text{%
\textrm{SYK}}},  \label{SYKzz}
\end{equation}%
where%
\begin{equation}
\left\langle g_{\beta }^{c}(\tau )\right\rangle _{\text{\textrm{SYK}}}=\int
dEdE^{\prime }\langle \varrho _{c}(E,E^{\prime })\rangle _{\mathrm{SYK}%
}e^{-(\beta +2ti)E}e^{-(\beta -2ti)E^{\prime }}.  \label{SYKc}
\end{equation}%
To calculate the first two items of Eq. (\ref{SYKzz}), we need to know the
spectral density of the SYK model, which has been derived by the method of
moments
\begin{equation}
\left\langle \varrho (E)\right\rangle _{\mathrm{SYK}}=\frac{1}{2\pi }\int
dte^{-iEt}\left\langle \text{Tr}(e^{iHt})\right\rangle _{\mathrm{SYK}}.
\end{equation}%
In this part, we are going to roughly estimate the dip and plateau time,
therefore, the spectral density can be approximated in a Gaussian type when $%
N$ is large \cite{Garcia-Garcia2016PRD,Garcia-Garcia2017PRD,Liu2017}%
\begin{equation}
\left\langle \varrho (E)\right\rangle _{\mathrm{SYK}}\simeq \sqrt{\frac{2}{%
\pi N}}d\exp \left( -\frac{2E^{2}}{N}\right) .
\end{equation}%
The partition function is
\begin{equation}
\left\langle Z(x)\right\rangle _{\mathrm{SYK}}\simeq d\exp \left( \frac{%
Nx^{2}}{8}\right) .  \label{gsykp}
\end{equation}%
With such equation, the decoherence time can be estimated by
\begin{equation}
\gamma \tau _{D}\geq \frac{1}{4\frac{\mathrm{d}^{2}}{\mathrm{d}\beta ^{2}}%
\ln \left[ \left\langle Z(\beta )\right\rangle _{\mathrm{SYK}}\right] }=%
\frac{1}{N}.
\end{equation}
\

Since the late time behavior of the \textrm{SYK} model is governed by GOE,
GUE, and GSE statistics according to the number of Majorana Fermions modulo $8$. For
simplicity, we first consider the connected part of the two-point
correlation function of GUE (i.e., $N$ mod $8=2$ or $6$) as
\begin{equation}
\langle \varrho _{c}^{(2)}(E,E^{\prime })\rangle _{\mathrm{SYK}}\simeq -\left(
\frac{\sin (2\pi r\left\langle \varrho (\omega )\right\rangle _{\mathrm{SYK}%
})}{\pi r}\right) ^{2},
\end{equation}%
with $r=E-E^{\prime }$ and $\omega =(E+E^{\prime })/2$ defined in above
subsection. Then
\begin{eqnarray}
\left\langle g_{\beta }^{c}(t)\right\rangle _{\text{\textrm{SYK}}} &\simeq
&\int dEdE^{\prime }\langle \varrho _{c}(E,E^{\prime })\rangle _{\mathrm{SYK}%
}e^{-(\beta +2it)E}e^{-(\beta -2it)E^{\prime }},  \notag \\
&=&-\int d\omega e^{-2\beta \omega }\int dr\left( \frac{\sin (2\pi
r\left\langle \varrho (\omega )\right\rangle _{\mathrm{SYK}})}{\pi r}\right)
^{2}e^{-irt},  \notag \\
&=&\left\{
\begin{array}{ll}
\frac{\sqrt{N}\left\langle Z(2\beta )\right\rangle _{\mathrm{SYK}}}{\sqrt{%
2\pi }d}t-2\left\langle Z(2\beta )\right\rangle _{\mathrm{SYK}}, & t\lesssim
2\sqrt{\frac{2\pi }{N}}d, \\
0, & t>2\sqrt{\frac{2\pi }{N}}d.%
\end{array}%
\right.   \label{gsykc}
\end{eqnarray}%
With Eqs. (\ref{gsykp}) and (\ref{gsykc}), the spectral form factor of the
SYK reads \
\begin{equation}
\left\langle g_{\beta }(t)\right\rangle _{\text{\textrm{SYK}}}\simeq \frac{%
\left\vert \left\langle Z(\beta +i2t)\right\rangle _{\mathrm{SYK}%
}\right\vert ^{2}}{\left\langle Z(\beta )\right\rangle _{\mathrm{SYK}}^{2}}%
+\left\{
\begin{array}{ll}
\frac{\sqrt{N}}{2\sqrt{2\pi }d}\left\langle g_{\beta }(\infty )\right\rangle
_{\text{\textrm{SYK}}}t, & t\lesssim 2\sqrt{\frac{2\pi }{N}}d, \\
\left\langle g_{\beta }(\infty )\right\rangle _{\text{\textrm{SYK}}} & t>2%
\sqrt{\frac{2\pi }{N}}d,%
\end{array}%
\right.   \label{gSYK}
\end{equation}%
where $\left\langle g_{\beta }(\infty )\right\rangle _{\text{\textrm{SYK}}}$
is the spectral form factor in the long time limit
\begin{equation}
\left\langle g_{\beta }(\infty )\right\rangle _{\text{\textrm{SYK}}}=\frac{%
2\left\langle Z(2\beta )\right\rangle _{\mathrm{SYK}}}{\left\langle Z(\beta
)\right\rangle _{\mathrm{SYK}}^{2}}\simeq \frac{2}{d}\exp \left( \frac{%
N\beta ^{2}}{4}\right) .
\end{equation}%
Note that when $N$ mod $8=0$, i.e., the GOE case, there is no degeneracy,
and the plateau height would be $\exp \left( N\beta ^{2}/4\right) /d$. From
Eq. (\ref{gSYK}), the plateau time is given by

\begin{equation}
t_{p}\simeq 2\sqrt{\frac{2\pi }{N}}d.  \label{plateau time}
\end{equation}%
For GOE ($N$ mod $8=0$) and GSE ($N$ mod $8=4$), the calculations would
become rather lengthy. Since we only aim to roughly estimate the time scale,
we still use the GUE, and modified the plateau time according to the
numerical results. For GSE, the plateau time is around $t_{p}\simeq \sqrt{%
2\pi /N}d$. Unlike the GUE and GSE, the ramp and plateau connect smoothly
for the GOE, so it is hard to strictly define the plateau time, for
simplicity, we still use Eq. (\ref{plateau time}) for estimation.

Before the dip time, the edges of the spectrum cannot be omitted, thus Eq. (%
\ref{gsykp}) is no longer applicable. Thus, we will replace it with $%
\left\langle Z(x)\right\rangle _{\mathrm{SYK}}\simeq x^{-3/2}$ \cite{Garcia-Garcia2016PRD,Garcia-Garcia2017PRD,Cotler2017}, and the first part of Eq. (\ref{gSYK}) is given by

\begin{equation}
\frac{\left\vert \left\langle Z(\beta +i2t)\right\rangle _{\mathrm{SYK}%
}\right\vert ^{2}}{\left\langle Z(\beta )\right\rangle _{\mathrm{SYK}}^{2}}%
\simeq \frac{\beta ^{3}}{(\beta ^{2}+c_{N}t^{2})^{3/2}},
\end{equation}%
with $c_{N}\simeq N/400$ fitted by numerical calculations. Then, the dip
time is roughly estimated as%
\begin{equation}
t_{d}\sim \left( \frac{\sqrt{\pi }\exp \left( -N\beta ^{2}/4\right) }{%
c_{N}^{3/2}\sqrt{2N}}\right) ^{1/4}\sqrt{d}\varpropto \sqrt{d}.
\label{dip time}
\end{equation}

\begin{figure*}[t]
\begin{center}
\includegraphics[width=0.85\linewidth]{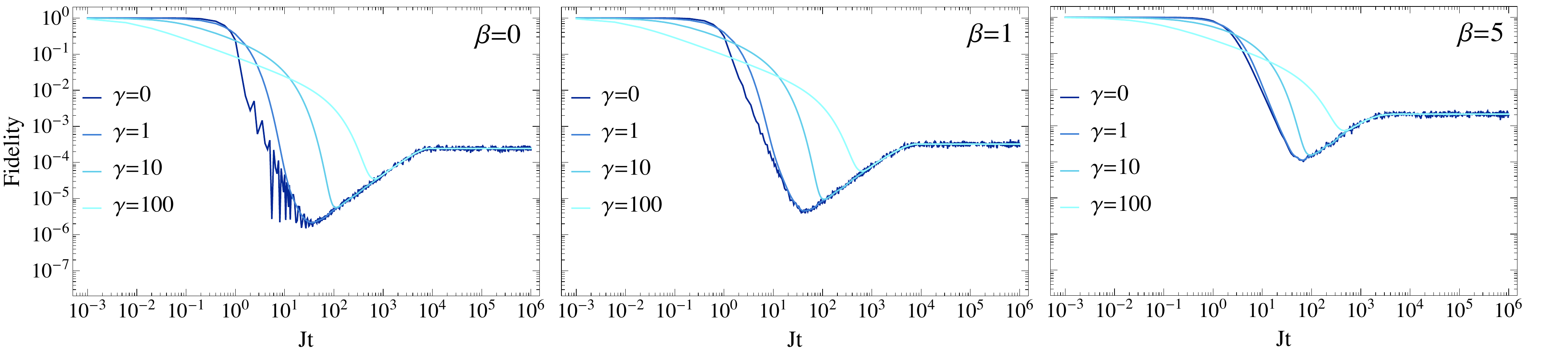}
\end{center}
\caption{\textbf{Fidelity of the stochastic SYK model for different
temperatures.} A log-log plot of the fidelity of SSYK model ($N=26$) under
different temperature. The variation of the temperature has a negligible effect on the dip and
plateau time.}
\label{SSYKfidelitybeta}
\end{figure*}

Although Eqs. (\ref{plateau time}) and (\ref{dip time}) are derived when $\beta$ is small,
they are still valid for low temperature for estimation,
just as shown in Fig. (\ref{SSYKfidelitybeta}).


\end{document}